\documentclass[12pt]{article}
\usepackage[margin=2cm]{geometry}
\usepackage{amsmath,amssymb,mathtools,graphicx,setspace}
\usepackage{cite}
\usepackage{slashed}
\usepackage{color}
\usepackage{hyperref}
\makeatother

\newcommand{\be}{\begin{equation}}
\newcommand{\bea}{\begin{eqnarray}}
\newcommand{\eea}{\end{eqnarray}}
\newcommand{\ba}{\begin{array}}
\newcommand{\ea}{\end{array}}
\newcommand{\ee}{\end{equation}}
\newcommand{\bes}{\begin{equation*}}
\newcommand{\beas}{\begin{eqnarray*}}
\newcommand{\eeas}{\end{eqnarray*}}
\newcommand{\bas}{\begin{array*}}
\newcommand{\eas}{\end{array*}}
\newcommand{\ees}{\end{equation*}}

\numberwithin{equation}{section}
\begin{document}
\onehalfspacing
\noindent

\begin{titlepage}
\vspace{10mm}

\vspace*{20mm}
\begin{center}
{\Large {\bf On Free Energy for Deformed  JT Gravity}\\
}

\vspace*{15mm}
\vspace*{1mm}
{Mohsen Alishahiha$^a$,  Amin Faraji Astaneh$^{b,c}$, Ghadir Jafari$^c$, Ali Naseh$^c$ and Behrad Taghavi$^c$}

 \vspace*{1cm}

{\it ${}^a$ School of Physics,
Institute for Research in Fundamental Sciences (IPM)\\
P.O. Box 19395-5531, Tehran, Iran
\\ \vspace{0.3cm}
 ${}^b$Department of Physics, Sharif University of Technology, Tehran 14588-89694, Iran
 \\ \vspace{0.3cm}
 ${}^c$  School of Particles and Accelerators, Institute for Research in Fundamental Sciences (IPM)\\
P.O. Box 19395-5531, Tehran, Iran
  }

 \vspace*{0.5cm}
{E-mails: {\tt alishah@ipm.ir, faraji@sharif.ir, ghjafari@ipm.ir, naseh@ipm.ir, btaghavi@ipm.ir}}

\vspace*{1cm}
\end{center}

\begin{abstract}
	
In this paper, we study a particular deformation of the Jackiw-Teitelboim gravity recently considered by Maxfield, Turiaci and  independently by Witten. We will compute the partition function of this model as well as its higher order correlators to all orders in genus expansion in the low temperature limit for small
perturbations. In this limit, the results match with those obtained from the Airy limit of a Hermitian random matrix ensemble. Using this result, we will also study the free energy of the model. One observes that although the annealed free energy has pathological behaviors, under certain assumptions, the quenched free energy evaluated by replica trick exhibits the desired properties at low temperatures.     

\end{abstract}

\end{titlepage}

\section{Introduction}
Unlike our general expectation{s from the} AdS/CFT correspondence, it was argued {by 
\cite{Saad:2019lba}} that Jackiw-Teitelboim (JT) gravity 
\cite{{Jackiw:1984je},{Teitelboim:1983ux}} is dual to a
 Hermitian random matrix model. As a result, the dual of JT gravity is not a specific quantum 
 theory but rather a random ensemble of quantum mechanical systems on the asymptotic 
  boundary. In this case, the Euclidean gravitational path integral should be thought of as 
  computing the ensemble average of the corresponding quantum mechanical systems.
 
 To be concrete, let us denote the partition function of a member of this ensemble by $Z(B)$, then one has 
\be
\langle Z(B)\rangle=\int_B Dg\; e^{-S},
\ee
that is the gravitational path integral over all geometries with a fixed boundary given by $B$. 
 Here, $\langle\cdots\rangle$ denotes the ensemble  average.

An important question to consider is whether such a partition function for the gravitational path integral factorizes when computed for geometries with disconnected boundaries. To be
precise, consider a gravitational path integral over geometries with boundary {$B^m=\cup_{i=1}^{m} B_i$}
\be
\langle Z(B)^m\rangle=\int_{\cup B_i} Dg\; e^{-S}.
\ee 
Then, it is not clear whether $\langle Z(B)^m\rangle=\langle Z(B)\rangle^m$. Indeed, recent progress in the 
context of black hole information paradox indicates that these two quantities 
are not equal due to contribution{s of Euclidean
replica wormholes} \cite{{Almheiri:2019qdq},{Penington:2019kki}}. In this case, a better 
definition of gravitational path integral with {multiple} disconnected boundaries may be given in 
terms of an ensemble average\cite{Engelhardt:2020qpv}
Actually, in the context of black hole information paradox, the {contributions from replica wormhole saddles were crucial for the fine-grained entropy to get the Page 
curve as is expected from unitarity} \cite{{Penington:2019npb},{Almheiri:2019psf},
{Almheiri:2019hni},{Almheiri:2019yqk},{Bousso:2019ykv}}.

Dealing with ensemble average{s,} it is important to note that in general a function of an 
averaged value is not equal to the average of the function, {\it i.e.} $f(\langle x\rangle)\neq \langle 
f(x)\rangle $, for a given function $f$.  In our context, this distinction becomes more significant 
when the contributions from the Euclidean wormholes are taken into account too. In particular, the 
free energy (which can be thought of as a function of partition function) may be computed in two 
different ways
\be
F_{\rm ann}=-T\;\ln \langle {Z(\beta)}\rangle ,\;\;\;\;\;\;\;\;\;F_{\rm que}=-T\;\langle {\ln Z(\beta)}
\rangle,
\ee
 where $\beta$ is the circumference of the boundary circle by which the  temperature of the system is given by $T=\beta^{-1}$ .  Note that since in the model we are considering the 
boundary is a deformed circle, the inverse temperature may be identified with the 
renormalized length 
of the boundary. We can see that due to the contributions of the Euclidean wormholes, the {\it 
annealed} free 
energy, $F_{\rm ann}$, is {not} equal to the {\it quenched} free energy, $F_{\rm que}$.

To further explore the role of Euclidean wormholes, the authors of \cite{Engelhardt:2020qpv}
have studied the contribution of Euclidean wormholes to the quenched free energy for
JT gravity and {a variant of the $CGHS$ model \cite{Callan:1992rs}, denoted as $\widehat{CGHS}$ model \cite{Afshar:2019axx},} by making {use of  the} replica trick in JT gravity and 
{$\widehat{CGHS}$} model. To do so, one should  consider the gravitational path integral
on $m$ copies of the boundary $B$, then it should be analytically continued 
to $m = 0$\cite{Engelhardt:2020qpv}
\be
F_{\rm que}=-T\;\langle {\ln Z(\beta)}\rangle=-T\lim_{m\rightarrow 0}\frac{1}{m}\left(\langle Z(\beta)^m\rangle -1\right){.}
\ee
From the above expression, it should be clear how the Euclidean wormholes contribute to the 
free energy. Actually, using the fact that
\be
\langle Z(\beta)^m\rangle=\langle Z(\beta)\rangle^m+ {\rm contributions\, of\, connected\, topologies{,}}  
\ee
one finds that if the dominant contribution is given by the disconnected topology, {\it i. e.} 
$\langle Z(\beta)^m\rangle\approx \langle Z(\beta)\rangle^m$, then the annealed and 
quenched free energies are the same. Otherwise, there is a big difference between annealed 
and quenched free energies.

Indeed, as was shown {in ref.}\cite{Engelhardt:2020qpv}, while the annealed free energy exhibits 
pathological behavior{s} at sufficiently low temperatures, the quenched free energy has a much 
better behavior at low temperatures, even though,  the quenched free energy is still 
not monotonic in 
this limit at least up to the finite order of truncation considered in their numerical computations. 
We will come back to this point later.

The aim of this article is to further explore the role of the replica wormholes by computing 
quenched free energy of a two dimensional gravity obtained by deformations of  JT
 gravity\cite{Mertens:2019tcm,Witten:2020ert,Witten:2020wvy,Maxfield:2020ale}. To do so, 
 we will first suggest a
 matrix model based on ``minimal strings'' as a possible dual
to the deform JT gravity which could  capture certain  non-perturbative information about 
deformed JT gravity. Using this dual description we will compute the quenched free energy
at low temperatures where we find that it exhibits the desired behavior.
 
This paper is organized as follows: In the next section, we shall study deformed JT gravity 
considered by Maxfield, Turiaci and  Witten.  We will compute the partition function of this model as well as its higher 
order correlators to all orders in genus expansion in the low temperature limit for small
 perturbations. We will also propose a matrix model dual based on the minimal string theory by which we could reproduce all previously obtained results.  In section 3, we will study 
 the annealed and quenched free energies of the deformed JT model using the 
 proposed matrix model where we see that non-perturbative effects play an important role. 
 Finally, the last section is devoted to discussions.


\section{Deformations of JT gravity}
In this section{,} we shall consider a 2-dimensional model obtained by a 
deformation of JT gravity studied in\cite{Mertens:2019tcm,Witten:2020ert,Witten:2020wvy,Maxfield:2020ale}\footnote{See also \cite{Momeni:2020tyt}
where the classical solution of the corresponding deformed gravity has also been studied}.  The most general 
action consisting of  two derivatives may be written as follows
\be\label{DJT}
I=-\frac{1}{2}\int d^2x \sqrt{g} (\phi R+W(\phi)),
\ee
where $W(\phi)$ is an arbitrary function of $\phi$. Setting $W(\phi)=2\phi+U(\phi)$, 
for parametrically small $U$, this model may be thought of as a perturbation of JT gravity.
Since the JT gravity is proposed to provide a gravity description for a random ensemble of  
quantum systems rather than a particular quantum system {defined} at the asymptotic region 
$\phi\rightarrow \infty$, it is natural to expect that 
the corresponding deformation of the JT gravity  given by the above action would also 
be dual to a random matrix ensemble with a different {spectral density} than 
the original JT gravity.

Indeed, it was argued in \cite{Witten:2020wvy,Maxfield:2020ale} that the deformed model, 
\eqref{DJT}, is dual to a hermitian matrix model that is obtained from the double-scaling limit of a 
matrix model with the measure $e^{-N \, { \rm Tr} V(H)}$, where $V$ is an arbitrary function of the 
Hamiltonian, $H$, of the dual quantum system.

The density of eigenvalues of the {ensemble}, $\rho(E)$, can be obtained from gravity partition 
function as follows
\be\label{DR}
\langle Z(\beta)\rangle =\int_{E_0}^\infty dE\; \rho(E)\,e^{-\beta E},
\ee
where $\beta$ is the circumference of the boundary circle and $E_0$ is the threshold energy. {Given 
a potential,  $W(\phi)$, one could in principle compute the gravitational path integral and then 
plug the result into the above equation and find the density of eigenvalues by which one may 
compute any other observables.} It was, however, noticed in \cite{Witten:2020wvy} that the 
correspondence between the potential and the spectral density depends on
 the renormalization scheme. To make the deformed model more tractable, the author of 
\cite{Witten:2020wvy} considered the following perturbation 
\bea\label{U}
U(\phi) = 2 \sum_{i=1}^{r} \varepsilon_{i}\hspace{.1cm} e^{-\alpha_{i }\phi}\;, \hspace{1cm} \pi < 
\alpha_{i}< 2\pi\,.
\eea 
For this particular perturbation, the gravitational path integral can be evaluated perturbatively 
and the result is given in terms of the Weil-Petersson volumes of moduli spaces of Riemann surfaces 
with conical singularities.

More precisely, to evaluate the gravitation path integral with  the perturbation \eqref{U}, 
following the procedure in JT gravity,  one first performs the integration over the scaler field 
$\phi$  resulting in a constraint saying  that
$R+2=0$ everywhere except at  $k$ conical singularities located at $x_i$
 with deficit angle $\alpha_i$ for $i=1, \dots, k$, in $k$-th level of perturbation. 
 
 The space of metrics  satisfying  this constraint, modulo diffeomorphisms, is  the moduli space of 
 Riemann surfaces of genus $g$ with $k$ {marked points} denoted by ${\cal M}_{g,k}$. In other words,
 ${\cal M}_{g,k}$ parametrizes  a family of hyperbolic Riemann surfaces with $k$ conical singularities of deficit angles $(\alpha_1, \dots,\alpha_k)$. 
 
\begin{figure}
\begin{center}
\includegraphics[scale=1.1]{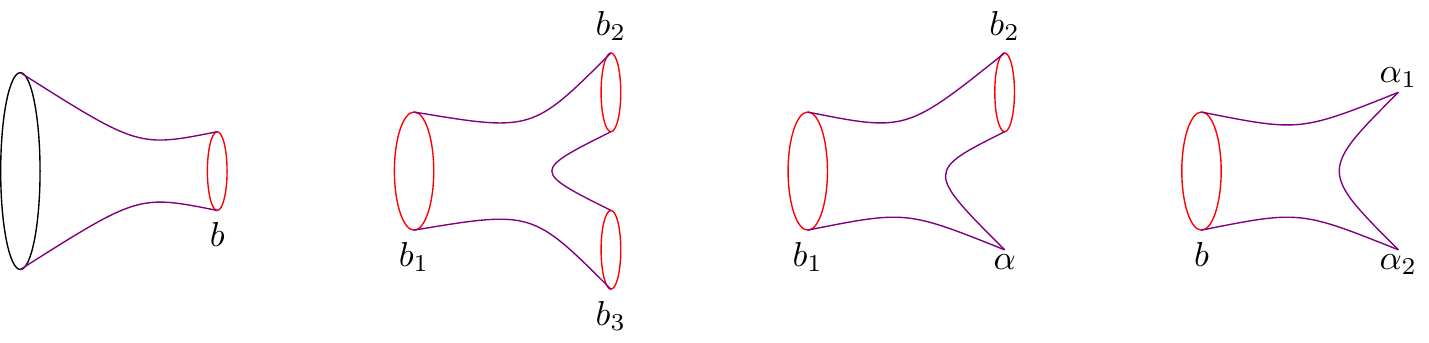}
\end{center}
\caption{Elementary building blocks to construct any hyperbolic Riemann 
surfaces with conical singularities. When the corresponding surface has no AdS boundary,
it can be constructed by  three-holed spheres (with or without conical singularities) and the 
trumpet is needed to have asymptotic AdS boundary.  The red
circles represent the geodesic boundaries while the black one is asymptotic AdS boundary. The 
geodesic boundaries' lengths and 
deficit angles are denoted by $b_i$ and $\alpha_i$, respectively. }  
\label{fig1}
\end{figure}

The same argument as that of JT gravity \cite{Saad:2019lba} can be applied in the present case
leading to the fact that the gravitational path  integral computes the volume of ${\cal M}_{g,k}$
with a modification due to a Schwarzian mode at asymptotic AdS boundary. The main technical
procedure utilized in \cite{Saad:2019lba} was that any complicated hyperbolic Riemann 
surfaces may be constructed by joining certain building blocks along suitable geodesics.
In the case of JT gravity there are two elementary  building blocks that are   a trumpet,
with an asymptotically AdS boundary and a geodesic boundary, and a three-holed sphere with 
three geodesic boundaries. In the present case one has two extra building blocks which can 
be obtained by three-holed sphere with replacing one or two geodesics with conical singularities
(see figure \ref{fig1}).

Following the earlier work of Mirzakhani\cite{M2007},  the Weil-Petersson volumes
we are interested in were studied in \cite{Tan:2006}. Indeed, the Weil-Petersson volumes of 
the moduli space of hyperbolic Riemann surfaces of genus $g$ with $m$ geodesic boundaries of 
lengths $(b_1, \dots, b_m)$ and $n$ conical singularities with deficit angles $(\alpha_1, \dots, \alpha_n)$ are given by \cite{Tan:2006}
\be\label{WP1}
V_{g,m,n}(b_1, \dots, b_m;\alpha_1, \dots, \alpha_n)=V_{g,m+n}(b_1, \dots,
b_m,i(2\pi-\alpha_1), \dots, i(2\pi-\alpha_{n}))\,.
\ee 
Here 
\be\label{WP0}
V_{g,q}(b_1,\dots,b_q)=\sum_{|d|+\ell=3g+q-3}\frac{(2\pi^2)^\ell}{2^{|d|}\ell !\,d_1!\dotsm
d_q!}b_1^{2d_1}\dotsm b_q^{2d_q}\int_{\overline{{\cal M}}_{g,q}}\kappa^\ell\psi_1^{d_1}
\dotsm \psi_q^{d_q},
\ee
is the Weil-Petersson volume of the moduli space of hyperbolic Riemann surfaces of genus $g$ with $q$ geodesic boundaries of 
lengths $(b_1, \dots, b_q)$\cite{M2}. In this expression  $\Psi_i$  is
the first Chern classes associated to the holes and $|d|=\sum_i d_i$. 
$\overline{{\cal M}}_{g,q}$ is the Deligne-Mumford 
compactification of ${\cal M}_{g,q}$ and $\kappa$ is  the first Mumford-Morita-Miller class on 
the $\overline{{\cal M}}_{g,q}$ (see \cite{Witten:2020wvy} for a quick review). 
 It is then clear that  the volume $V_{g,q}(b_1, \dots, b_q)$ 
is a  polynomial in $b_{i}^2$ of degree $(3g-3+q)$. 

Now, we have all the ingredients  to compute the gravitational path integral over surfaces 
with $m$
boundaries. As we already mentioned{,} to evaluated the corresponding integrals one needs
to consider two  contributions. The first one comes from the  Schwarzian modes associated 
to the asymptotic AdS boundary and the second part comes from the Weil-Petersson volumes.
In the present case{,} and with the deformation \eqref{U}{,} the contribution of the Schwarzian 
modes are given by \cite{Witten:2020wvy}
\be
Z_{\text{Disk}}(\beta) =\frac{e^{\frac{\pi^2}{\beta}}}{4\sqrt{\pi}\beta^{3/2}},
\hspace{1cm}Z_{D(\alpha)}
(\beta)=\frac{e^{\frac{(2\pi-\alpha)^2}{4\beta}}}{2\sqrt{\pi\beta}},\hspace{1cm}
Z_{\text{trumpet}}(b,\beta) = \frac{e^{-\frac{b^2}{4\beta}}}{2\sqrt{\pi\beta}}\,.
\ee
Here we have  assumed that the asymptotic boundary is a deformed circle with a renormalized 
length $\beta$. With this notation, one may compute the gravitational path integral for 
a connected geometry with $m$ boundaries  which, in turns, corresponds to evaluating 
the connected part of the $m$-correlator 
\bea\label{Pconn}
\langle Z(\beta)^m\rangle_{\rm conn} = \sum _{g=0}^{\infty} e^{-S_{0}\left(2g+m-2\right)}
\hspace{.1cm}Z_{g,m}(\beta),
\eea
where $S_0$  is the (renormalized) ground state entropy and (see also \cite{Witten:2020wvy})
\bea\label{Zs}
 Z_{0,1}(\beta) \!&\!\!=\!\!&\!  Z_{\text{Disk}}(\beta)
 \cr &&\cr &&\!\! +\sum_{i=1}^{r}\varepsilon_{i}Z_{D(\alpha_i)}(\beta)
 +\sum_{n=1}^\infty\sum_{i_1, \dots, i_{n}=1}^{r} \frac{1}{n!}\hspace{.1cm} \varepsilon_{i_1} \dotsm
\varepsilon_{i_n}
 \int_{0}^{\infty} db\hspace{.1cm}b \hspace{.1cm}Z_{\text{trumpet}}(b,\beta)\hspace{.1cm} V_{0,1;n}
 (b;\alpha_{i_1}, \dots, \alpha_{i_n}),
\cr &&\cr  
Z_{0,2}(\beta) \!&\!=\!&\! 
\int _{0}^{\infty}db\hspace{.1cm} b\hspace{.1cm} Z^2_{\text{trumpet}}(b,\beta)
\cr &&\cr
&&\!\!+\sum_{n=1}^\infty\sum_{i_1, \dots, i_n=1}^{r} \frac{1}{n!}\hspace{.1cm} \varepsilon_{i_1}\dotsm
\varepsilon_{i_n} \int_{0}^{\infty} \bigg(\prod_{j=1}^2 db_j\hspace{.1cm}b_j 
\hspace{.1cm}Z_{\text{trumpet}}(b_j,\beta)
\bigg)\hspace{.1cm} V_{0,2;n}(b_1, b_2;\alpha_{i_1}, \dots, \alpha_{i_n}),
\cr &&\cr  
Z_{g,m}(\beta) \!&\!=\!&\! \int_{0}^{\infty} \bigg(\prod_{j=1}^m db_j\hspace{.1cm}b_j 
Z_{\text{trumpet}}(b_j,\beta)\bigg)\hspace{.1cm} V_{g,m}(b_1, \dots, b_m)
\\ &&\!\!
+\sum_{n=1}^\infty \sum_{i_1, \dots, i_{n}=1}^{r} \frac{1}{n!}\varepsilon_{i_1}\dotsm \varepsilon_{i_{n}}
\int_{0}^{\infty} \bigg(\prod_{j=1}^m db_j b_j \hspace{.1cm}Z_{\text{trumpet}}(b_j,\beta)\bigg)
V_{g,m;n}(b_1, \dots, b_m;\alpha_{i_1}, \dots, \alpha_{i_n}).\nonumber
\eea
Here the first line of each expression represents  the contribution of JT gravity whereas the
second line is  the contribution of the deformation of JT gravity. See the figure 2 for 
visual inspiration of
how to write these expressions. 
\begin{figure}
\begin{center}
\includegraphics[scale=1]{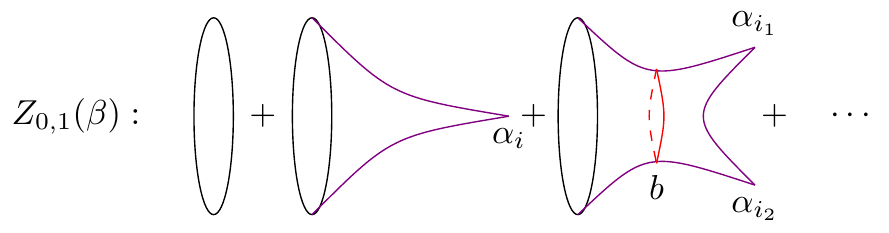}
\includegraphics[scale=1]{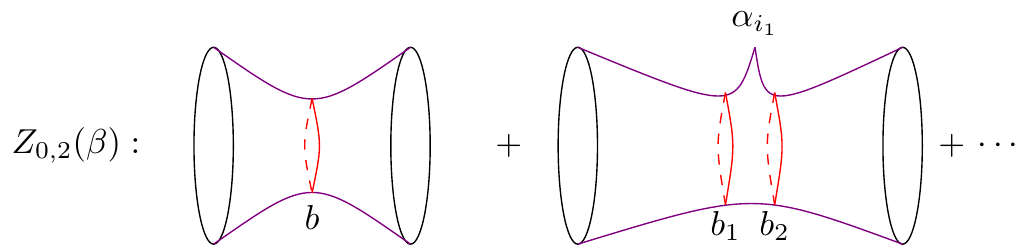}
\end{center}
\caption{A visual  representation of how to compute, for example,
$Z_{0,1}(\beta)$ and $Z_{0,2}(\beta)$. The red circles represent the ``suitable'' geodesics.}
\label{fig2}
\end{figure}
Note that the factor of $\frac{1}{n!}$ comes from the fact 
that the location of conical singularities are  indistinguishable\footnote{When conical singularities 
come with different ``flavors" (\textit{i.e.} different deficit angles and strength), the symmetry factor 
is given by $\prod_{i=1}^{N_f} \frac{1}{n_i !}$ where $n_i$ is the multiplicity of each flavor and 
$N_f$ is the total number of flavors.}  \cite{Witten:2020wvy}.
Since the Weil-Petersson volumes are polynomials in $b_{i}$, one can perform the integration over
 $b$ in the expressions of $Z_{g,m}$  to find  $Z_{g,m}$ as a function of $\beta$. 

\subsection{Gravity correlation  functions}

In this subsection we would like to compute the gravitational path integral using 
the above procedure. To proceed we shall first compute the disc partition 
function from which one may read the  density of eigenvalues of the dual 
double-scaled random matrix model (assuming there is such  a dual matrix model). 

The first few geometries 
contributing to the disc partition function are depicted in 
figure \ref{fig2}. The corresponding partition function is\cite{Witten:2020wvy}
\bea\label{Z0W}
&& \langle Z(\beta)\rangle_{g=0}=e^{S_0}Z_{0,1}(\beta)
=\frac{e^{S_0}}{4\sqrt{\pi}\beta^{\frac{3}{2}}}\bigg(e^{\frac{\pi^2}{\beta}}
+2\beta \sum_{i=1}^r \varepsilon_i e^{\frac{(2\pi-\alpha_i)^2}{4\beta}}+2{(\sum_i^r
\varepsilon_i)^2 \beta^2}+\frac{4}{3}(\sum_i^r\varepsilon_i)^3 \beta^3\cr &&\cr
&&\hspace{47mm}+\frac{4}{3}\bigg(4\pi^2\sum_i^r\varepsilon_i -3
\sum_i^r\varepsilon_i(2\pi-\alpha_i)^2\bigg)(\sum_i^r\varepsilon_i)^2\beta^2\bigg)+\cdots,
\eea
which at low temperatures (large $\beta$) and small perturbations can be written in
the following suggestive form
\bea\label{Zg0}
&& \langle Z(\beta)\rangle_{g=0}= \frac{e^{S_0}}{4\sqrt{\pi}\beta^{\frac{3}{2}}}
\left(1
+\beta (2 \sum_i^r \varepsilon_i) 
+\frac{1}{2!}{(2\sum_{i=1}^r\varepsilon_i)^2 \beta^2}+\frac{1}{3!}(2\sum_i^r\varepsilon_i)^3 
\beta^3+\cdots\right)\cr &&\cr
&&\hspace{2.1cm}\approx \frac{e^{S_0}}{4\sqrt{\pi}
\beta^{\frac{3}{2}}}e^{U(0)\beta}+\cdots,
\eea
where $U(0)=2\sum_{i=1}^r\varepsilon_i$.  Assuming that there is a dual double-scaled matrix
model, its corresponding density of eigenvalues can be obtained by making use  of an 
inverse Laplace
transformation  \cite{Witten:2020wvy}
\be\label{RW}
\rho(E)\approx e^{S_0}\frac{\sqrt{E+U(0)}}{2\pi},
\ee
which is consistent with the general near threshold behavior of a hermitian matrix model
if one identifies the threshold energy by $E_0=-U(0)$. Taking other terms into
account one gets the same expression, though with a threshold energy with  higher order 
correction\cite{Witten:2020wvy}.

It is also found useful  to compute the contributions of higher genus terms to the partition
 function. 
Of course,  in general this would be a very hard task. Nonetheless, for small enough 
perturbations, one could still make a progress in evaluating the corresponding quantity. 
Indeed, in this case  one needs to
compute the partition function given by (see also \cite{Okuyama:2019xbv} for JT gravity) 
\be\label{E1}
\langle Z(\beta)\rangle=e^{S_0}  Z_{0,1}(\beta)+
\sum_{g=1}^\infty e^{-S_0(2g-1)}Z_{g,1}(\beta),
\ee
where
\be\label{E2}
Z_{g,1}(\beta) = \frac{1}{2\sqrt{\pi\beta}}\int_{0}^{\infty} db\hspace{.1cm}b\; 
e^{-\frac{b^2}{4\beta}}\bigg(V_{g,1}(b)
+\sum_{n=1}^\infty \sum_{i_1, \dots, i_{n}=1}^{r} \frac{1}{n!}\varepsilon_{i_1}\dotsm \varepsilon_{i_{n}}
V_{g,1;n}(b;\alpha_{i_1}, \dots, \alpha_{i_n})\bigg).
\ee
Here using  the equations \eqref{WP1} and  \eqref{WP0} one has
\bea
&&V_{g,1}(b)=\sum_{d+\ell=3g-2}\frac{(2\pi^2)^\ell}{2^d d!\ell!}b^{2d}
\int_{\overline{{\cal M}}_{g,1}}\kappa^\ell \psi_1^d,\\ &&\cr
&&V_{g,1;n}(b,\alpha_{i_1}\cdots,\alpha_{i_n})
=\hspace{-5mm}
\sum_{q+d+\ell=3g+n-2}\hspace{-3mm}\frac{(-1)^q(2\pi^2)^\ell}{2^{d+q} d!\ell! q_1{!}
\cdots q_n!}
b^{2d}\prod_{j=1}^n(2\pi-\alpha_{i_j})^{2q_j}\int_{\overline{{\cal M}}_{g,1+n}}\kappa^\ell \psi_1^d\psi_2^{q_1}
\cdots \psi_{1+n}^{q_n}\,.\nonumber
\eea
with $q=\sum_{i=1}^n q_i$. Although it might be difficult to compute these volumes exactly, 
it is a relatively easy task to 
compute the leading order term in the low temperature limit by choosing the maximum value for 
$d$ in the above expressions {\it -- i.e.} $d=3g-2$ for the first volume and $d=3g+n-2$ for the 
second one. Doing so, one gets\footnote{Note that we have used the fact that 
 $\int_{\overline{{\cal M}}_{g,k+1}}\psi^{3g+k-2}=\frac{1}{24^g g!}$, $k=0,1\cdots$ 
 (see for example \cite{MB}).}
\be
\int_{0}^{\infty} db\hspace{.1cm}b\; 
e^{-\frac{b^2}{4\beta}}V_{g,1}(b)=\frac{1}{2\beta}\;\frac{(\beta^3/3)^g}{g!}+\cdots,\;\;\;\;\;\;\;\;
\int_{0}^{\infty} db\,b\; 
e^{-\frac{b^2}{4\beta}}V_{g,1,n}(b)=(2\beta)^{n-1}\;\frac{(\beta^3/3)^g}{g!}+\cdots\;.
\ee
It is now straightforward  to perform summations over $n$ and genus $g$
appearing in the equations \eqref{E1} and \eqref{E2} to arrive at
\be\label{Pc1}
\langle Z(\beta)\rangle \simeq\frac{e^{S_0}}{4\sqrt{\pi}\beta^{\frac{3}{2}}} 
e^{\frac{\beta^3}{3}e^{-2S_0}}e^{U(0)\beta}+\cdots\,,
\ee
that should be compared with \eqref{Zg0}. One observes that the summation over $n$ results
in $e^{U(0)\beta}$ factor, while the contribution of 
summing over all genus yields to the factor of ${\rm Exp}[\frac{1}{3}\beta^3e^{-2S_0}]$.
Since we are working in {the} large $\beta$ limit, summing over all genus 
may not be convergent. Note, however that since we are working in the semi-classical 
gravity limit, and in order to be able to trust the gravity description, one should work in the 
large $S_0$ limit too. Therefore, to have control over the model, we will be working in the 
double scaling limit; in which both $\beta$ and  $e^{S_0}$ are large while 
keeping $\beta^{\frac{3}{2}}e^{-S_0}$ fixed.

It is also interesting to  study the connected part of $m$-correlators, $\langle Z(\beta)\rangle_{\rm
conn}$, using the equation \eqref{Pconn}. To do so, one needs to evaluate $Z_{g,m}(\beta)$ given 
in the equation \eqref{Zs} which in turn requires to evaluate the Weil-Petersson volumes for 
arbitrary $m$\footnote{We note that asymptotic behavior of Weil-Petersson volume has been
studied in \cite{{Zograf},{Kimura:2020zke}} which could be used to compute partition function
at large genus limit.}. Of course, since we are interested in the low temperature 
limit of the $m$-point 
function, in the expression of Weil-Petersson volumes given in the equation\eqref{WP0}, one may
set $\ell=0$ 
\be
V_{g,m}(b_1, \dots, b_m)=\sum_{|d|=3g+m-3}\frac{1}{2^{|d|}\,d_1!\dotsm
d_m!}b_1^{2d_1}\dotsm b_m^{2d_m}\int_{\overline{{\cal M}}_{g,m}}\psi_1^{d_1}
\dotsm \psi_m^{d_m}.
\ee
Similarly, for the case of $n$ conical singularities, the powers of 
$(2\pi-\alpha_i)$ factors should also be set to zero in addition to $\ell$. 
Therefore, in this case the 
corresponding volumes  do not depend on  $\alpha_i$'s. More explicitly, one has 
\be
V_{g,m,n}(b_1, \dots, b_m;\alpha_1, \dots, \alpha_n)=\hspace{-5mm}
\sum_{|d|=3g+m+n-3}\frac{1}{2^{|d|}\,d_1!\cdots
d_m!}b_1^{2d_1}\cdots b_m^{2d_m}\int_{\overline{{\cal M}}_{g,m+n}}\psi_1^{d_1}
\cdots \psi_m^{d_m}.
\ee
Plugging the above expressions into the equation \eqref{Zs} and performing the integrals 
over $b_I$'s and, moreover, taking into account the following identity (see for example \cite{Do})
\be
\int_{\overline{{\cal M}}_{g,m+1}}\psi_1^{d_1}
\cdots \psi_m^{d_m}=\sum_{k=1}^m\;\;\; \int_{\overline{{\cal M}}_{g,m}}\psi_1^{d_1}\cdots
\psi_k^{d_k-1}
\cdots \psi_m^{d_m},
\ee
one gets 
\bea
Z_{g,m}(\beta) \!&\!=\!&\!\left(\frac{\beta}{\pi}\right)^{\frac{m}{2}}\bigg(\sum_{|d|=3g+m-3}
(2\beta)^{3g+m-3}\int_{\overline{{\cal M}}_{g,m}}\psi_1^{d_1}
\cdots \psi_m^{d_m}\\ &&\hspace{2cm}+ \sum_{i_1,\cdots,i_{n}=1}^{r} \frac{2^n}{n!}
\varepsilon_{i_1}\cdots \varepsilon_{i_{n}}\;m^n\;\beta^n\sum_{|d|=3g+m-3}(2\beta)^{3g
+m-3}\int_{\overline{{\cal M}}_{g,m}}\psi_1^{d_1}
\cdots \psi_m^{d_m}\bigg).\nonumber
\eea
Inserting this expression in the equation \eqref{Pconn} one arrives at
\be\label{Pconn1}
\langle Z(\beta)^m\rangle_{\rm conn}=\left(\frac{\beta e^{-\frac{2S_0}{3}}}{\pi}\right)^{\frac{m}{2}}
e^{mU(0)\beta}
\sum_{g=0}^\infty\sum_{|d|=3g+m-3}\prod_{i=1}^mt_i^{d_i}
\int_{\overline{{\cal M}}_{g,m}}\psi_1^{d_1}
\cdots \psi_m^{d_m},
\ee
where $t_i=2\beta e^{-\frac{2S_0}{3}}$ for all $i$'s. Actually, the summations over all genus and 
all possible partitions of $d_i$ appearing in this expression have been evaluated in \cite{OKO}
that may be written as follows
\bea\label{OO}
\sum_{g=0}^\infty\sum_{|d|=3g+m-3}\prod_{i=1}^mt_i^{d_i}
\int_{\overline{{\cal M}}_{g,m}}\psi_1^{d_1}
\cdots \psi_m^{d_m}\!\!&\!\!=\!\!&\!\!\left(\frac{2\pi}{t_i}\right)^{\frac{m}{2}}\hspace{-0.5cm}\sum_{\rm all\; possible\; k-partitions\; of\; m}\hspace{-1cm}(-1)^{k-1}
{\cal E}_k(m_1,\cdots,m_k)\cr &&\cr &&\hspace{-5.4cm}=\left(\frac{2\pi}{t_i}\right)^{\frac{m}{2}}
\bigg({\cal E}_1(m)-m\; {\cal E}_2(m-1,1)+\cdots+(1)^{m-1}(m-1)!\;{\cal E}_m(1,\cdots,1)\bigg),
\eea
where ${\cal E}_k$ is
\be\label{Zmk}
{\cal E}_k(m_1, \dots, m_k)=\left( \frac{e^{S_0}}{\sqrt{\pi}} \right)^k
 \;\frac{e^{\frac{\sum_im_i^3}{3}\beta^3 e^{-2S_0}}}{\beta^{\frac{k}{2}}
 \prod_{i=1}^km_i^{\frac{1}{2}}}\;
 \int_{-\infty}^{0}\prod_{j=1}^kdx_j\;e^{\sum_{i=1}^k(x_i+x_{i-1})m_i \beta}
 e^{-\sum_{i=1}^k\frac{(x_i-x_{i-1})^2}{4m_i\beta e^{-2S_0}}}.
\ee
Note that in the summation over all 
possible $k$-partitions  of $m$, one should also take into account the symmetric factor for
 each term as indicated in the second line of \eqref{OO}. Plugging this expression into the 
equation \eqref{Pconn1} one finds
\be\label{ZM1}
\langle Z(\beta)^m\rangle_{\rm conn}=\sum_{\rm all\; possible\; k-partitions\; of\; m}(-1)^{k-1}
{\cal Z}(m_1,\dots,m_k),
\ee
with 
 \be
{\cal Z}(m_1,\dots,m_k)=e^{m U(0)\beta }{\cal E}_k(m_1,\dots,m_k).
\ee
Note that in this summation the proper symmetric factor for
 each term, as mentioned,  should be also taken into account.

\subsection{Matrix model description  } 

In this section, assuming that the deformed JT gravity \eqref{DJT} is dual to a double-scaled random matrix model, we would like to study a possible dual matrix model that
reproduces the desired results of the gravity computations. 

It is worth mentioning that the main technical reason of having a duality between a double-scaled  matrix model 
and JT gravity shown in \cite{Saad:2019lba} was that upon a Laplace 
transformation  Mirzakhani's recursion relation \cite{M4} maps to the recursion relation of a
double-scaled matrix model\cite{Eynard:2007fi}.
 The corresponding recursion relation could  fix all correlators in all orders in 
 genus expansion in terms of  the inverse Laplace transformation 
of disc partition function (the leading order density of eigenvalues). 
So that, to all orders in $e^{-S_0} $ expansion the arbitrary correlation functions, 
$\langle Z(\beta_1)\cdots Z(\beta_n)\rangle$, of JT gravity coincides with that of a
double-scaled random matrix ensemble. Since the model we have been considering in this 
paper is a deformation of JT gravity one would naturally expect 
that a similar  procedure could also work for the deformed JT duality\cite{Maxfield:2020ale}.

It was also argued that the double-scaled matrix model relevant for JT gravity has a 
certain  similarity 
with minimal string theory   \cite{Saad:2019lba}. The connection with minimal string 
theory and its possible relation with the old version of matrix model of 2-dimensional 
gravity has also been studied in \cite{Okuyama:2019xbv} and 
\cite{{Johnson:2019eik},{Johnson:2020heh},{Johnson:2020exp}}. An advantage of working with
minimal string theories  is  that one may have a better 
control of non-perturbative effects of gravity in this context.

Following \cite{Johnson:2019eik} in this subsection we shall explore the above  procedure
for deformed JT gravity. To proceed, we note that in this context  the partition function of desired
matrix integrals may be viewed as the partition  function of a one dimensional quantum system  with the following Hamiltonian 
\cite{Gross:1989aw,Banks:1989df,Douglas:1989dd}
\be\label{HH}
H=-e^{-2S_0}\frac{\partial^2}{\partial x^2}+u(x),
\ee
where $u(x)$ is a potential to be determined for the model  of interest. Indeed, 
the corresponding potential satisfies a non-linear ordinary differential equation known as 
 string equation\cite{Gelfand:1975rn}\footnote{In general one could consider the
 case in which the right hand side of this equation is non-zero, given by $e^{-S_0}\Gamma^2$
 with $\Gamma$ being a constant \cite{{Johnson:2019eik},{Johnson:2020heh}}.  It would be 
 interesting to study this model too.}
\bea\label{KdV}
 (u-E_0) \mathcal{R}^2 - \frac{e^{-2S_0}}{2} \mathcal{R} \mathcal{R}'' + \frac{e^{-2S_0}}{4} 
(\mathcal{R}')^2 = 0,
\eea
where
\be 
 \mathcal {R}\equiv \sum_{k=0}^{\infty}t_k R_k[u]+x.
\ee
Here  ${R}_k[u]$ is a $k$-th polynomial  of  $u(x)$ and its derivatives with respect to $x$
defined in \cite{Gelfand:1975rn}. In particular in our notation 
\be
R_0[u]=1,\;\;\;\;\;\;\;R_1[u]=u(x),\;\;\;\;\;\;\;\;{R}_2[u] = u(x)^2 - \frac{1}{3}e^{-2S_0} u''(x).
\ee
To find the coefficients $t_k$'s for the deformed JT gravity one may  use that the fact that 
for $x=0$ setting  $R_k=E_0^k$ then ${\cal R}=0$ reduces to an algebraic
equation from which one could find the exact threshold energy $E_0$.

To find the corresponding algebraic equation we note that at low temperatures 
the leading order density of eigenvalues is  given by  $\rho_0(E)\sim \sqrt{E-E_0}$, which 
in turns, fixes the general form of the disc partition function for large $\beta$ as 
follows \cite{Witten:2020wvy}
\bea
&& \langle Z(\beta)\rangle_{g=0}
=\frac{e^{S_0}e^{-\beta E_0}}{4\sqrt{\pi}\beta^{\frac{3}{2}}}
\bigg\{e^{\beta E_0}
\bigg(e^{\frac{\pi^2}{\beta}}
+2\beta \sum_{i=1}^r \varepsilon_i e^{\frac{(2\pi-\alpha_i)^2}{4\beta}}+2{(\sum_i^r
\varepsilon_i)^2 \beta^2}+\frac{4}{3}(\sum_i^r\varepsilon_i)^3 \beta^3\cr &&\cr
&&\hspace{42mm}+\frac{4}{3}\bigg(4\pi^2\sum_i^r\varepsilon_i -3
\sum_i^r\varepsilon_i(2\pi-\alpha_i)^2\bigg)(\sum_i^r\varepsilon_i)^2\beta^2\bigg)+\cdots
\bigg\},
\eea
with a constraint that  coefficients of the positive powers of $\beta$ in the brackets should 
vanish. In particular from  vanishing of the coefficient of $\beta$ one gets
\cite{{Witten:2020wvy},{Maxfield:2020ale}}
\be
\frac{\sqrt{E_0}}{2\pi}I_1(2\pi\sqrt{E_0})+\sum_{i=1}^r\varepsilon_i I_0((2\pi-\alpha_i)\sqrt{E_0})=0,
\ee 
where $I_0$ and $I_1$ are  modified Bessel functions. Using the series  expansion of the modified Bessel
functions, one can read  the coefficients $t_k$'s as follows   
\be\label{TK}
 t_k=\frac{1}{2}\frac{\pi^{2k-2}}{k!(k-1)!}+ \sum_{i=1}^{r} 
\varepsilon_{i} \frac{(2 \pi-\alpha_{i})^{2k}}{2^{2k} (k!)^2},\;\;\;\;\;\;\;\;
t_0=\sum_{i=1}^{r} 
\varepsilon_{i}
\ee

In general, using the string equation, one could find the potential  associated to the 
Hamiltonian of the quantum system \eqref{HH}.  Then, one could in principle 
find the  spectrum of the Hamiltonian from which one  may evaluate the spectral density 
non-perturbatively. 
 
Of course, in general it is a very hard task. We note, however, that at low temperatures
one could find the potential perturbatively. Indeed,  this is exactly the Airy limit in which 
the corresponding  matrix model is controlled by the universal behavior of the edge of
 the classical density of  eigenvalues\cite{Forrester:1993vtx}. 
 
The potential we are interested in can be obtained by the solution of $\mathcal{R}=0$ which 
at leading order results in
$
u(x)=-2C x +E_0$, 
with
\be
C=1-\pi U'(0)-\frac{1}{4}U''(0),\;\;\;\;\;\;\;\; E_0=-U(0)+\frac{1}{2}\pi^2 U^2(0)+
\pi U(0)U'(0)+\frac{1}{ 4}U(0)U''(0)\,,
\ee
where
\be
U'(0)=-2\sum_{i=1}^r\varepsilon_i\alpha_i,\;\;\;\;\;\;\;\;\;\;\;U''(0)=2\sum_{i=1}^r\varepsilon_i
\alpha_i^2.
\ee
It is then clear that the Hamiltonian gets a shift with $E_0$. Therefore one arrives at the following 
equation for the wave function 
\be
\left[- \frac{e^{-2S_0}}{2C} \frac{\partial^2}{\partial x^2} -x \right] \psi(E,x) = \frac{E-E_0}{2C}
 \, \psi(E,x)\,,
\ee
which could be solved to find
\bea
 \psi(E,x) = e^{\frac{2 S_0}{3}}(2C)^{\frac{1}{3}} \operatorname{Ai}(\xi),\;\;\;\;\;\;{\rm with}\;\; \xi =
  - e^{\frac{2 S_0}{3}}(2C)^{\frac{1}{3}} \left(x + \frac{E-E_0}{2C}\right).
\eea
Using this wave function, the corresponding density of eigenvalues,
\be\label{rho}
\rho_{\rm Airy}(E)=e^{\frac{4 S_0}{3}}(2C)^{\frac{2}{3}}\int_{-\infty}^0 dx\; {\rm Ai}^2\left[ -
 e^{\frac{2 S_0}{3}}(2C)^{\frac{1}{3}} \left(x + \frac{E-E_0}{2C}\right)\right],
\ee
reads
\bea\label{DAiry}
\rho_{\text{Airy}}(E)=e^{\frac{2 S_0}{3}}(2C)^{\frac{1}{3}}\left(\text{Ai}'^2(
\eta)-\eta \text{Ai}^2(\eta)\right),\;\;\;\;\;{\rm with}\;\;\;
\eta= -  e^{\frac{2 S_0}{3}}(2C)^{\frac{1}{3}}\; \frac{ E-E_0}{2C}.
 \eea
In particular, at leading order in $e^{-S_0}$ it is
 \be\label{DAL}
\rho_{\text{Airy}}(E)\approx \frac{e^{S_0}}{\pi}\;\sqrt{E-E_0},\;\;\;\;\;\;\;\;\;\;\;\;\;{\rm for }\;\;E\geq E_0,
\ee 
 reproducing  \eqref{RW}. It is worth noting that although the above leading order 
 expression for energy density is valid for $E\geq E_0$, the full expression \eqref{DAiry}  has a non-zero support on the entire real axis. In fact, in what follows, in order to compare the results with 
those obtained from the geometrical computations we will take all integral over the entire real axis,
 though in general one  should be careful about the spectrum below the threshold energy
\cite{{Johnson:2019eik},{Johnson:2020exp}}. We will come back to this point in the next 
section where we shall compute the free energy.  

 Plugging the energy  density  \eqref{DAiry} into the equation \eqref{DR} one gets (see below 
 for more details)
\bea\label{Pm6} 
\langle Z(\beta)\rangle 
=\frac{e^{S_0}}{2\sqrt{\pi}\beta^{\frac{3}{2}} }e^{\frac{1}{3}\beta^3 e^{-2S_0}} 
e^{- \beta E_0 },
\eea
which is the same as that obtained from gravity computations \eqref{Pc1}.

In what follows we will compute 
$\langle Z(\beta)^m\rangle_{\rm conn}$ to all orders in genus expansion for arbitrary $m$
in the low  temperature limit using the Airy wave function of the dual matrix model 
considered above.  To proceed, we note that the connected $m$ point function consists of 
several terms  given by \cite{Okuyama:2018aij}({see also\cite{Johnson:2020mwi,Okuyama:2020ncd}})
\bea\label{PC}
\langle Z(\beta)^m\rangle_{\rm conn}&=&{\rm Tr}
\left(e^{-m\beta H}P\right)
-m {\rm Tr}\left(e^{-(m-1)\beta H}Pe^{-\beta H}P\right)+\cdots \cr &&+(-1)^{m-1}(m-1)!\;
{\rm Tr}\left(e^{-\beta H}Pe^{-\beta H}P\cdots P e^{-\beta H}P\right)
\eea
where $P$ is the projection operator
\be
P=\int_{-\infty}^0 dx\;|x\rangle\langle x|.
\ee
To evaluate this connected $m$ point function, one generally needs to compute
the following quantity 
\be
{\cal Z}(m_1, \dots, m_k)={\rm Tr}\left(e^{-m_1\beta H}Pe^{-m_2\beta H}P\cdots P e^{-m_k\beta 
H}P\right)
\ee
with the condition {$\sum\limits_{i=1}^k m_i =m$}. By making use of  the wave function $H |\psi
\rangle =E|\psi\rangle$ and the fact {that $\int dE |\psi\rangle\langle \psi|=1$, one could rewrite the 
traces in terms of integrals in order to arrive at} 
\be
{\cal Z}(m_1, \dots, m_k)=\int_{-\infty}^{0}\prod_{j=1}^kdx_j
\prod_{i=1}^k\int_{-\infty}^{\infty} dE_i\;\; e^{-m_i\beta E_i}\;
\psi(E_i,x_{i-1})\psi^*(E_i,x_i)\,,
\ee
where 
\bea
 \psi(E_i,x_j) = e^{\frac{2 S_0}{3}}2^{\frac{1}{3}} \operatorname{Ai}
 \left[ - e^{\frac{2 S_0}{3}}2^{\frac{1}{3}} \left(x_j + \frac{E_i-E_0}{2}\right)\right].
\eea
Here we have used a notation in which $x_0=x_k$. Plugging this wave function into the 
above equation and using a proper change of the coordinates,   one gets
\bea
 &&{\cal Z}(m_1, \dots, m_k)=\left(e^{\frac{2S_0}{3}}2^{\frac{4}{3}} \right)^k\;e^{-m\beta E_0}
  \int_{-\infty}^{0}\prod_{j=1}^k dx_j\\
&&\hspace{1cm}\times \prod_{i=1}^k
  \int_{-\infty}^{\infty} dE_i\;\; e^{
e^{\frac{-2 S_0}{3}}2^{\frac{2}{3}}m_i\beta E_i}{\rm Ai}
\left(E_i- e^{\frac{2 S_0}{3}}2^{\frac{1}{3}} x_{i-1} \right)
{\rm Ai}\left(E_i- e^{\frac{2 S_0}{3}}2^{\frac{1}{3}} x_i\right).\nonumber
\eea
By making use of the following identity 
\be\label{ID}
\int_{-\infty}^{\infty} e^{xy}\text{Ai}(x+a)\text{Ai}(x+b)\, dx=\frac{1}{2\sqrt{\pi y}}\exp\left(\frac{y^3}
{12}-\frac{a+b}{2}y-\frac{(a-b)^2}{4y}\right)\,,
\ee
one may perform the integration over $E_i$'s  to get
\be
{\cal Z}(m_1, \dots, m_k)= e^{-m\beta E_0}\;
{\cal E}_k(m_1, \dots, m_k),
\ee
where ${\cal E}_k(m_1, \dots, m_k)$ is given by \eqref{Zmk}. Therefore, one gets 
the same expression for connected $m$ corrlators as given in the equation \eqref{ZM1}.
Clearly, for $k=1$ this reduces to that of \eqref{Pm6}. 

It is worth noting that  the connected $m$ point function at low temperatures decays  
as $\beta^{-3/2}$ while the partition function of fully disconnected topologies with $m$ 
boundaries  goes as $\beta^{-3m/2}$. It  indicates that at low temperatures the physics is
dominated by the connected topologies.


\section{Free energy for Deformed JT gravity}

In the previous section, we have considered a particular deformation of JT gravity whose 
dual theory is shown to be a Hermitian random matrix model
\cite{Witten:2020wvy,Maxfield:2020ale}. We have also computed the partition function
of this model at low temperatures to all orders in genus expansion for small perturbations. 
We have observed that the resultant partition function (in the low temperature limit) is 
in agreement with that obtained from the Airy limit of a Hermitian matrix model.

Having found the partition function{,} it is then natural to proceed to study other relevant 
quantities. One of the most fundamental {quantities} one may study is the free energy.  
Thus, in what follows, we will study free energy following the procedure  considered 
in \cite{Engelhardt:2020qpv}. Of course, since the computations are very similar to that of 
the mentioned paper in the first  part of this section, we will be very brief and the reader 
is referred  to the paper for more details.

\begin{figure}[h]
\center\includegraphics[scale=.45]{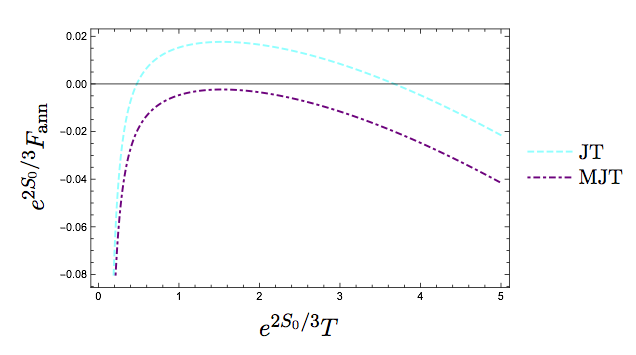}
\caption{The annealed free energy for $S_0=7$ and $U(0)=1/50$. Apart {from} the divergence
for 
$\beta \gg 1$, the local maximum is still present even by the inclusion of higher genus {corrections}
(doubly non-perturbative effects).}\label{f1}	
\end{figure}

An immediate
way to compute the free energy is to consider the logarithm of connected partition function 
that computes {the} annealed free energy 
\be\label{Fann}
F_{\text{ann}}=- T \log \langle Z(\beta)\rangle = - T \log \langle Z(\beta)\rangle\,,
\ee
which{, using equation \eqref{Pc1}, is given by}
\be
F_{\text{ann}} = -T{S_0}-T\log\frac{T^{\frac{3}{2}}}{4\sqrt{\pi}}-\frac{1}{3T^2}e^{-2S_0}-{U(0)}\,.
\ee
Using the full expression for {the} partition function, given in equation \eqref{Pconn}, one can 
numerically evaluate the annealed free energy and the result of this computation is depicted in 
figure \ref{f1}. To compare this result with that of JT gravity, we have also plotted 
the corresponding free energy for this model. 

It is worth mentioning  that this figure cannot be trusted for high {temperatures}.   Indeed, we  
should emphasize that the main lesson one can learn from this figure is that 
the annealed free energy diverges and also exhibits a maximum at low temperatures. Therefore, the annealed free 
energy is not a useful quantity to be considered for the system since it is both non-monotonic and divergent.
Indeed, as it is well-known in condensed matter physics literature, a better quantity one may 
define is the  quenched free energy which we shall study it in what follows.

To compute the quenched free energy for JT gravity, the authors of  \cite{Engelhardt:2020qpv} 
have used a certain replica trick\footnote{The quenched free energy for Gaussian matrix model
has been studied in\cite{Okuyama:2020mhl} using a direct computation of ensemble 
average of the logarithm of partition function.}. 
In the present case, we will consider the same method to 
compute the quench free energy. To proceed, one should consider the gravitational path integral
on $m$ copies of the boundary $B$ and then the analytical continuation
to $m = 0$ should be performed
\be\label{Fque}
F_{\rm que}=-T\langle \log Z(\beta)\rangle=-T\lim_{m\rightarrow 0}\frac{1}{m}\left(\langle Z(\beta)^m\rangle-1\right)\,.
\ee
Although the above expression is a concrete statement, in practice it is very hard to implement. 
 The main point 
in this calculation is that since the Weil-Peterson volumes are only well-defined for integer 
$m$, it is not clear how to extend their definition to the case of non-integer $m$.
 Therefore, the analytical continuation 
for general $m$, and in particular the limit $m\rightarrow 0$, might not be well-imposed.

The way  the authors of  \cite{Engelhardt:2020qpv} proposed to circumvent  this difficulty was to 
define a ``{\it truncated free energy}'' by evaluating the gravitational path integral over geometries
which include  only topologies that connect up to M boundaries. More precisely, one may 
define $\langle Z(\beta)^m\rangle_{M}$ as follows
\be
\sum_{m=0}^\infty \frac{t^m}{m!}\langle Z(\beta)^m\rangle_{M}=
{\rm exp}\left(\sum_{m=1}^M \frac{t^m}{m!}\langle Z(\beta)^m\rangle_{\rm conn}\right),
\ee
and then the truncated free energy is defined by 
\be
\bar{F}_{M}=-T\lim_{m\rightarrow 0}\frac{1}{m}\left(\langle Z(\beta)^m\rangle_M-1\right)\,.
\ee
For more detail the reader is refereed to  \cite{Engelhardt:2020qpv}.
Following this paper, let us considered the following analytic continuation 
of  $m$ factorial 
\bea
 m! = \int_{0}^{\infty} dt\hspace{.1cm}e^{-t}\hspace{.1cm}t^{m},\;\;\;\;\;\;\;\;\;\;
 \frac{1}{m!} = \frac{1}{2\pi i} \oint_{C\equiv\text{Hankel countour}} dz
 \hspace{.1cm}e^{z} z^{-(m+1)}.
\eea
\begin{figure}[h]
\hspace{.1cm}
\includegraphics[scale=.44]{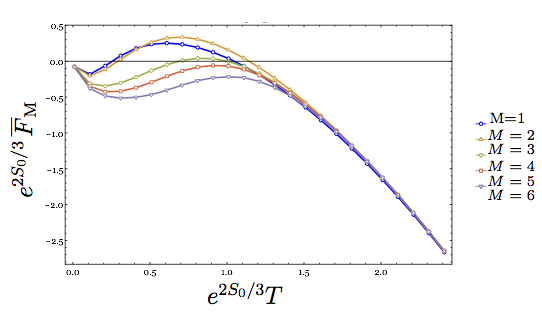}
\includegraphics[scale=.44]{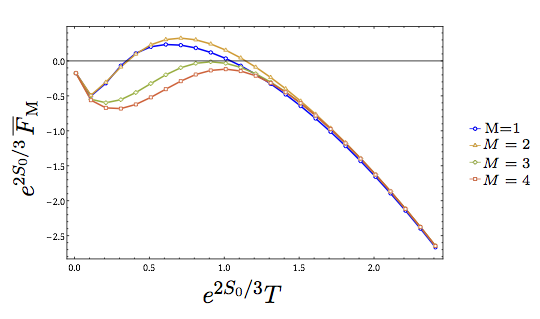}
\caption{{\it Left}: The (first order perturbation) quenched free energy $\overline{F}_{M}$ for 
$S_0=7$, $\epsilon=1/100$ and $\alpha=3\pi/2$ using topologies with genus up to one and 
various $M =1,2,3,4,5,6$. {\it right}: The (first order perturbation) quenched free energy 
$\overline{F}_{M}$ for $S_0=7$, $\epsilon=1/100$ and $\alpha=3\pi/2$ using topologies with genus up to two  and various $M =1,2,3,4$.}
\label{fig4}	
\end{figure}
Then, the truncated free energy is found to be \cite{Engelhardt:2020qpv}
 \bea
\overline{F}_{M} = - \text{T}\hspace{.1cm} \text{Re} \int \bigg(
\prod _{k=1}^{M-1}d\mu(t_k,z_k)\bigg)\frac{1}{M!}\sum_{j_1=0}^{1}...\sum_{j_{M-1}=0}^{M-1} \log \mathcal{A}_{j_{1}...j_{M-1}}^{(M)}(\{t_{k},z_{k}\})
\eea
where
\bea
&&d\mu(t_k,z_k) = \int_{0}^{\infty} dt_{k}\oint dz_k\hspace{.5mm}\frac{e^{z-t}}{2\pi i z},
\cr \nonumber\\
&& \mathcal{A}_{j_{1}...j_{M-1}}^{(M)}(\{t_{k},z_{k}\}) = \mathcal{P}_{\text{conn},1}(\beta)+
\sum _{k=2}^{M}e^{\frac{2\pi i}{k}j_{k-1}}\bigg(\frac{\langle Z(\beta)^k\rangle_{\rm conn}}{k!
\hspace{.1cm} z_{k-1}}\bigg)^{\frac{1}{k}} t_{k-1}. 
\eea
Therefore, the quenched free energy is defined as
\be
F_{\rm que}=\lim_{M\rightarrow\infty} \overline{F}_M.
\ee
It is worth mentioning that the resultant free energy depends on the choice of 
contours, $C_k$, of which the integration over $z_k$'s is performed (needed for the 
analytic continuation of the $\Gamma(m+1) \equiv m!$).
Although, using this definition of analytic continuation makes it possible to compute 
the quenched free energy numerically, in practice it is very time consuming  (if not impossible) 
to evaluate large $M$ limit of the truncated free energy. The numerical results for few $M$'s are
depicted in the figure 4.

As it is clear from figure 4, the replica wormholes have significantly changed the low 
temperature behavior of the free energy, although it still has its undesirable feature of being 
non-monotonic! One reason for this odd behavior could be that the quenched free energy, 
by definition, is supposed to be computed in the large $M$ limit. 
Therefore, even though the plots get smoothed already 
in small $M$'s, we should not expect to have the correct behavior for small $M$'s.

Of course, one could still doubt about the analytic continuation of $m$ and the way the 
replica trick is used in the case. This issue has been extensively explored in 
\cite{Engelhardt:2020qpv} where the results were interpreted as 
an evidence for the necessity of replica symmetry breaking.  
We note, however, that since the observation
is  based on the numerical computations, the results may not be conclusive enough. 
It would be interesting if one could have a better understanding of low temperature behavior
of the free energy and in particular if one could do some analytic computations. This is, 
indeed, the subject of the next subsection.


\subsection{Free energy at low temperatures}

As we have seen in the previous section{,} the deformed JT gravity at low temperatures 
could be described by a matrix model in the  Airy limit, in which the corresponding 
matrix model is controlled by the universal behavior of the edge of the classical density of 
eigenvalues. Using this observation{,} in what follows we would like 
to compute the quenched free energy of {this} model at low temperatures by making use 
of the  Airy wave function.

To proceed, one needs to compute the gravitational path integral on  geometries with 
$m$ boundaries, $\langle Z(\beta)^m\rangle$.
In general, the gravitational path integral involves a sum over all topologies connecting $m$ 
boundaries and
the resultant quantity should be interpreted as being dual to an ensemble average
(at least in two dimensions). In particular, for $m=1$, this is actually ensemble average  
of partition function  studied in the previous section.

 The obtained partition function may be  used  to 
compute the annealed free energy that is essentially the same as that in  \eqref{Fann}, 
showing pathological properties  for free energy at the  low temperatures. 
This, in turn,  would indicate the necessity of using replica trick to include 
the connected topologies.  

For  $m$ disconnected boundaries , there are different topologies contributing to gravitational 
path integral. If the gravitational path integral was computed for the fully disconnected topology, 
one had 
$ \langle Z(\beta)^m\rangle=\langle Z(\beta)\rangle^m$. We note, however, that in the 
gravitational theory we are considering, this is not the case.  Therefore, we will have 
to sum over all topologies. The sum starts from the fully disconnected geometry and will end up 
with the fully connected one. Schematically, one may write 
\be\label{ZM}
\langle Z(\beta)^m\rangle=\langle Z(\beta)\rangle^m+\langle Z(\beta)\rangle^{m-2}
\langle Z(\beta)^2\rangle_{\rm conn}+\cdots+\langle Z(\beta)^m
\rangle_{\rm conn}\,.
\ee
The first term is just the $m$-th power of the partition function we have computed in the 
previous section that decays as $\beta^{-3m/2}$ for large $\beta$. 
The fully connected term is also computed in the previous section where 
one observed  that at leading order it decays as $\beta^{-3/2}$ at  low temperatures.  

For fixed $m\geq1$ the connected topologies dominate at low temperatures while 
at high temperatures, the fully disconnected one plays an important role. 
The temperature in which 
the role of connected and disconnected topologies gets exchanged is given in terms of 
the extremal entropy $e^{S_0}$. In fact, the situation is very similar to that of the island formula
for fine grained entropy of evaporating black holes \cite{{Almheiri:2019qdq},{Penington:2019kki}}.
Indeed, in our study,  $\beta^{-1}$ plays the role of the matter degrees of freedom in that context.

We note, however,  that since we are interested in using replica trick to evaluate free energy 
in which one needs to analytically continue $m$ for non-integer numbers and in particular taking 
$m\ll 1$ limit, the situation might be more involved.  At this point, all we would like to emphasize 
is that taking the $m\rightarrow 0$ is tricky and one should keep track of how to implement  this limit.

In the previous section, we have computed the connected part of the gravitational path integral 
with $m$ boundaries at low temperatures. In principle, one could plug the result into the 
equation \eqref{Fque} to find the quenched free energy. Doing so, one gets
\be
F_{que}\approx-T \lim_{m\rightarrow 0}\frac{1}{m}\left(Z(\beta^m)_{\rm conn}-1\right)\sim
\lim_{m\rightarrow 0}
\left(-
\frac{T^{\frac{5}{2}} e^{S_0}}{\sqrt{\pi}m^{\frac{5}{2}}}
e^{\frac{m}{3}\beta^3e^{-2S_0}} e^{-m\beta E_0}+\frac{T}{m}
\right){.}
\ee
The above expression seems divergent and leads to an ill-defined quantity. Keeping in mind the 
above discussions on analytic continuation, our understanding of this undesired result is that 
the naive replica trick and $m\rightarrow 0$ limit  we have used are not precise enough.

To further explore this point, we will follow the procedure which has been
 recently considered by \cite{Johnson:2020mwi} in the remainder of this section. To begin, let us 
 note that the spectral density in the Airy limit, \eqref{DAiry}, is non-zero at the threshold energy:
 \be
 \rho_{\rm Airy}(E_0)= \frac{e^{\frac{2S_0}{3}}2^{\frac{1}{3}}}{3^{\frac{2}{3}} \Gamma 
 \left(\frac{1}{3}\right)^2}\,.
 \ee
Note that non-zero energy density at threshold energy should be generated by 
non-perturbative quantum effects. In particular, looking at the leading order 
{(g=0)} expression for the spectral density, \eqref{DAL}, one observes that it vanishes at 
the threshold energy.

 Having a finite and non-zero density, it is natural to study the thermal properties of this 
 model around the threshold energy (specially if one wants to work in the low temperature 
 approximation).  Being at a thermal state, 
 an appropriate way to identify the state is to give its spectral density which is defined in
 \eqref{rho}. This expression can be written in the following form
 \be
\rho_{\rm Airy}(\mu,E)=e^{\frac{2 S_0}{3}}2^{\frac{1}{3}}\int^{ 2^{\frac{1}{3}} e^{\frac{2 {S_0}}{3}}
\mu}_0 dx\; {\rm Ai}^2\left[x -
 e^{\frac{2 S_0}{3}}2^{-\frac{2}{3}} \left( {E-E_0}\right)\right],
\ee
where $\mu>0$ is a cutoff whose significance will be become clear latter. It is worth 
mentioning that $\rho_{\rm Airy}(E_0)=\lim_{\mu\rightarrow \infty}\rho_{\rm Airy}(\mu,E_0)$. 
Of course, in what follows we shall  consider the  opposite limit, $\mu\rightarrow 0$, in which 
 one has $\rho_{\rm Airy}(\mu,E)\sim \mu$. The reason for taking small $\mu$ limit is that in order 
 to maintain  the validity of our approximation for expression of the  wave function
 around the threshold energy the range of $x$ should be small.
  
As we have already mentioned in the previous section, although the energy density 
defined in the Airy limit  (given in terms of the Airy function) may be defined along all real axis,
in order to compute the partition function one needs to truncate the energy 
for $E\geq E_0$. In fact, motivated by that of the  JT gravity \cite{Saad:2019lba} 
one would expect that  there are  non-perturbative instabilities which could be connected to 
the fact  that  there are non-perturbative contributions to the spectral density for $E< E_0$. 
In the context of JT gravity, a remedy to avoid the instability was proposed in
\cite{{Johnson:2019eik},{Johnson:2020exp}} that essentially enforces us to work 
with the  Airy function with the spectrum truncated at $E=0$. Following 
this observation in what follows we will consider the case where the spectrum is 
truncated for $E\geq E_0$.

It is important to mention that restraining ourselves to $E\geq E_0$  limit,  the truncated 
spectrum would change the leading order behavior of the partition function significantly.
To see this, let us compute the  partition function around the threshold energy as follows
 \be
\langle Z(\beta)\rangle= \int_{E_0}^{\infty} dE\;
e^{-\beta E} \rho_{\rm Airy}(\mu,E_0)=\frac{\rho_{\rm Airy}(\mu,E_0)}{\beta}e^{-\beta E_0} .
\ee
 On the other hand, since we are at low energies,
one should also take $\mu\rightarrow 0$ limit
 \be\label{Dmu}
\rho_{\rm Airy}(\mu,E_0) \approx \frac{2^{\frac{2}{3}} e^{\frac{4 {S_0}}{3}}}{3^{\frac{4}{3}} 
\Gamma \left(\frac{2}{3}\right)^2}  \;\mu,\;\;\;\;\;\;\;\;{\rm for}\;\;\mu\approx 0\,.
\ee
so that 
\be
\langle Z(\beta)\rangle\approx  \frac{2^{\frac{2}{3}} }{3^{\frac{4}{3}} \Gamma
 \left(\frac{2}{3}\right)^2} \frac{e^{\frac{4 {S_0}}{3}}\mu}{\beta}e^{-\beta E_0} .
\ee
Note that, unlike the equation \eqref{Zg0}, the leading order of the above partition
function decays as $\beta^{-1}$ at low temperatures.  It is also useful to evaluate  
${\cal Z}(m_1,\dots,m_k)$ in this limit:
\be
{\cal Z}(m_1,\dots,m_k)=
 \frac{\rho^k_{\rm Airy}(\mu,E_0)}{\beta^k\prod_{i=1}^k m_i}\;
e^{-m\beta E_0}.
\ee
Then, it is straightforward to compute the gravitational path integral over geometries with 
$m$ boundaries \eqref{ZM}
\bea\label{ZM1}
&&\langle Z(\beta)^m\rangle =\bigg\{\bigg(\frac{\rho^m_{\rm Airy}(\mu,E_0)}{\beta^m}\bigg) 
+\frac{m(m-1)}{2}\bigg(\frac{\rho^{m-1}_{\rm Airy}(\mu,E_0)}{2\beta^{m-1}} 
- \frac{\rho^{m}_{\rm Airy}(\mu,E_0)}{\beta^{m}}\bigg)
+\cdots\\ &&\cr
&&\hspace{1cm}+m\bigg(\frac{\rho^2_{\rm Airy}(\mu,E_0)}{(m-1) \beta^2 }
-(m-1)  \frac{\rho^3_{\rm Airy}(\mu,E_0)}{\beta^3(m-2)}\;
+\cdots
+(-1)^{m-2}(m-2)! \frac{\rho^{m}_{\rm Airy}(\mu,E_0)}{\beta^{m}} \bigg)\cr &&\cr
&&\hspace{1cm} +\bigg(\frac{\rho_{\rm Airy}(\mu,E_0)}{m \beta }\;
-m  \frac{\rho^2_{\rm Airy}(\mu,E_0)}{\beta^2(m-1)}\;
+\cdots
+(-1)^{m-1}(m-1)! \frac{\rho^m_{\rm Airy}(\mu,E_0)}{\beta^m}\;
\bigg)\bigg\}e^{-m\beta E_0}. \nonumber
\eea 
Here, the first term corresponds to the fully disconnected piece while the last one 
comes from the fully connected term. The other terms are a mixture of the connected and 
disconnected components. In particular, the two other terms that are explicitly 
written in the above equation, are associated with the following terms 
\be
\langle Z(\beta)\rangle^{m-2}
\langle Z(\beta)^2\rangle_{\rm conn}\;,\;\;\;\;\;\;\;\;\;\;\;\;
\langle Z(\beta)\rangle
\langle Z(\beta)^{m-1}\rangle_{\rm conn}\,.
\ee
Now,  the aim is to use the replica trick to find the free energy. To do so, we note that 
from equation \eqref{ZM1} one finds that taking $m\rightarrow 0$ in the 
following  equation is rather tricky 
\be
\langle \log Z(\beta)\rangle=\lim_{m\rightarrow 0}\frac{1}{m}(\langle Z(\beta)^m\rangle-1)\,.
\ee
Indeed, since we are interested in the low temperature limit (in which $\beta$ is very large),
 one essentially  needs to take two limits: $\beta\rightarrow \infty$ and $m\rightarrow 0$ and, in 
 fact,  the order in which the limits are taken is very crucial. In order to find a finite result,
 taking into account \eqref{Dmu},  it turns out that one needs to consider the large 
 $\beta$ limit at first\footnote{Actually our main insight to take the limit of $\beta\rightarrow \infty$ first comes from the details of calculations. In particular one would, generally, expect that at low temperatures the main contribution to the correlation functions comes from the connected part. 
 The ordering we have considered here is consistent with this expectation. It is worth noting 
 had we considered the 
 limit of $m\rightarrow 0$ first, we would have lost the main contribution 
 of connected part at low temperatures.}, then  one finds
\be\label{Zrp}
\langle Z(\beta)^m\rangle \approx \frac{\rho_{\rm Airy}(\mu,E_0)}{m \beta }e^{-m\beta E_0}.
\ee
Moreover, the cutoff $\mu$ should be set to $m^2$.  Doing so one arrives at
\be
\langle \log Z(\beta)\rangle=\frac{2^{\frac{2}{3}} e^{\frac{4 {S_0}}{3}}}{3^{\frac{4}{3}} \Gamma 
\left(\frac{2}{3}\right)^2}\frac{1}{\beta}-\frac{1}{m}\,.
\ee
Interestingly enough the main contribution to the quenched free energy that leads to a 
finite result comes from $\langle Z(m\beta)\rangle$ term that is the most symmetric one.
Namely, we have a boundary with the length $m\beta$. This might indicate that the replica 
symmetry is respected in this procedure of finding quenched free energy. Of course, there is
a subtlety to this conclusion as one has non-zero spectral density at zero temperature.
It would be interesting to understand this point better. 

Another issue worth exploring is the $\frac{1}{m}$ divergent term appearing in the 
expression of the ensemble average of the logarithm of partition function.
 Actually, the origin of this 
divergent term should be traced back to the way we took large $\beta$ limit and 
the way we have thrown away different terms. In other words, it has to do with the regularization 
of the partition function. Indeed, there could be a term that might vanish at large $\beta$ limit 
though remains finite in the $m\rightarrow 0$ limit. 
Therefore, in general, one could expect to have an extra term in 
the expression of \eqref{Zrp} that although it might be small in low the temperature limit, it
 could grow to be of order one as we are taking the  $m\rightarrow 0$ limit. This extra 
 piece could regularize the $\frac{1}{m}$ term appearing in the 
replica trick leading to an extra constant in the expectation value of the 
logarithm of partition function. More precisely, one should have
\be
\langle \log Z(\beta)\rangle=\frac{2^{\frac{2}{3}} e^{\frac{4{S_0}}{3}}}{3^{\frac{4}{3}} \Gamma 
\left(\frac{2}{3}\right)^2}\frac{1}{\beta}+Z_0.
\ee 
This is also consistent with the fact that the model in this limit can be obtained 
by double scaling limit of a matrix model as explored in \cite{Johnson:2020mwi}.
The extra constant term may be fixed by evaluating entropy of the system from logarithm 
of the partition function
\be
S(\beta)=(1-\beta \partial_\beta) \langle \log Z(\beta)\rangle=Z_0+\frac{2^{\frac{7}{3}} e^{\frac{4
 {S_0}}{3}}}{3^{\frac{4}{3}} \Gamma 
\left(\frac{2}{3}\right)^2}\frac{1}{\beta},
\ee 
which results in $Z_0=S_0$ that is the extremal entropy. Therefore, the quenched free 
energy is found to be
\be
F_{\rm que}=-T\langle \log Z(\beta)\rangle=-\frac{2^{\frac{2}{3}} 
e^{\frac{4 {S_0}}{3}}}{3^{\frac{4}{3}} \Gamma 
\left(\frac{2}{3}\right)^2}\; T^2-S_0T.
\ee 
The above expression exhibits the desired properties for a free energy at low temperatures. It is 
important to remind ourselves that this result is valid for extremely low temperature limit.
It is interesting to note that at leading {order,} the threshold energy does not appear in this 
expression and moreover {the} leading term is proportional to $(T{e^{\frac{2 S_0}{3}}})^2$
as expected from replica wormholes \cite{Engelhardt:2020qpv}.


\section{Discussions}

In this paper we have studied certain features of  a particular deformation of JT gravity 
considered in \cite{{Witten:2020wvy},{Maxfield:2020ale}}. This particular deformation, being
in an exponential form, keeps the model tractable. Indeed, the effect of the deformation 
could be studied perturbatively using  a geometric description for the
gravitational path integral in terms of the  Weil-Petersson volumes with a number of  
punctures.  Such volumes have been already studied in the literature following the  
novel work of Mirzakhani. 

Based on this description for the corresponding deformation of the JT gravity, 
we have evaluated the partition function 
as well as its higher order correlators to all orders of genus expansion in
the low temperature limit and for small perturbations.  

The results have been compared with those obtained from a 
matrix model in the Airy limit, in which the corresponding matrix model is controlled by the 
universal behavior of the edge of the classical density of eigenvalues, that are in complete 
agreement. This observation is in favor of the conjecture that this particular deformation 
of JT gravity has a dual description in terms of a Hermitian random matrix model.
 We have also seen that this particular deformation leads to a shift in the partition function 
given by ${\rm Exp}[-\beta E_0]$ where at leading order the threshold energy is $E_0=-U(0)$. 
Our results show that inclusion of higher genus contribution and higher order perturbation 
leave this pattern unchanged, though one should consider the  threshold energy with 
higher order corrections.

Using this result, we have also studied the free energy of the model. The dual theory, being 
a random ensemble of quantum mechanical systems rather than a specific quantum theory,
requires to evaluate the physical quantities while taking  an ensemble average.  It is then 
important to distinguish between the logarithm of ensemble average of the partition function 
and the ensemble average of the logarithm of the partition function. While the first 
one computes the annealed partition function, the latter should be interpreted as the quenched 
free energy.

Following \cite{Engelhardt:2020qpv}, we have computed the annealed and quenched free 
energies for the deformed JT gravity using numerical methods to compute the 
 Weil-Petersson volumes. 
We have observed that the annealed free energy has pathological behaviors 
at low temperatures. More precisely, it results in a non-monotonic free energy which 
could yield to a negative specific heat. Therefore, the quenched free energy is a better 
quantity to be considered.

In order to compute the quenched free energy, one may utilize a replica trick by which 
the replica wormholes may contribute to make the obtained free energy well behaved.
Indeed, adding the replica wormholes will smooth the behavior of the free energy at low
temperatures, though it is not enough to fully remove the  pathological behavior. Actually,
the situation is exactly the same as that of JT gravity itself without deformation, studied 
in \cite{Engelhardt:2020qpv} where it was argued that this undesired behavior 
should be related to the analytic continuation to $m=0$. Therefore, in order to have a clear 
understanding, one needs to study this analytic continuation in more details.

In order to explore this point, we have utilized the fact that
 the deformed JT gravity at low temperatures could be described by a matrix model in the Airy 
 limit. Thus, we have used an analytic computation to evaluate the free energy at low 
 temperatures by making use of the Airy wave function.
 
 Our analytic considerations were based on two crucial points. First of all, due to non-perturbative 
 effects the energy density at threshold energy is a finite non-zero number. Therefore, we have 
 studied the quenched free energy around the threshold energy. Moreover, we have truncated 
 the energy spectrum to consider just $E\geq E_0$, even though the Airy density is non-zero for all energies. 
 As a result of this truncation, the obtained partition function at low temperatures has a leading 
order that decays as $\beta^{-1}$. This should be compared with the leading order term 
of the partition function obtained from a fully geometric computation given in terms of the
Weil-Petersson volumes that decays as $\beta^{-3/2}$.

In this approximation, we have computed the gravitational path integral for geometries with 
$m$ boundaries by which one could compute the free energy using the replica trick. 
We have seen that it was 
very crucial to take the $m\rightarrow 0$ limit while we are interested in the low temperature limits. 
Going through the detailed computations, we have found that the main 
contribution to the quenched free energy comes from the fully connected topologies. 
Indeed,
it is even more interesting that among all of the connected topologies, the most dominant 
contribution is
 coming from the most symmetric one, \textit{i.e.} $\langle Z(m\beta)\rangle$. This means that we 
 are dealing with a geometry with just 
one boundary with length $m\beta$. Therefore, one might conclude that 
the analytic continuation and the  limit of $m\rightarrow 0$  in the replica trick preserves 
the replica symmetry.  And the resultant free energy exhibits the desired properties 
at low temperatures. 

Of course, there are still a few points which have not been fully understood yet. 
In particular, we have seen that the ensemble average of the logarithm of the partition 
function needs to be regularized in order to get a finite 
result that required to produce the extremal entropy. We should admit that our argument 
in favor of this point is not rigorous enough.

Another point worth mentioning is that the numerical results based on the truncated 
free energy obtained from the Weil-Petersson volumes seem unable to fully reproduce 
the correct free energy even though if we compute it for large enough $M$. 
The point is that partition 
function obtained from the Weil-Petersson volumes (even at all orders in the genus expansion)
does not contain non-perturbative effects which seem important to find the desired result. 
In other words, it seems that the geometric partition function based on the 
Weil-Petersson volumes is not be capable to produce the quenched free energy correctly.    
 
It would be interesting to further explore this point which in turn could result in
a deeper understanding of the analytic continuation to $m=0$.


\section*{Acknowledgements}

 We  would like to thank Hamid Reza Afshar for useful discussions.


\end{document}